\documentclass[lettersize,journal]{IEEEtran}
\usepackage[nolist]{acronym}
\usepackage{amsmath,amsfonts}
\usepackage{algorithmic}
\usepackage{algorithm}
\usepackage{array}
\usepackage[caption=false,font=normalsize,labelfont=sf,textfont=sf]{subfig}
\usepackage{textcomp}
\usepackage{stfloats}
\usepackage{url}
\usepackage{verbatim}
\usepackage{graphicx}
\usepackage{subcaption} 
\usepackage[export]{adjustbox}
\usepackage{wrapfig}
\usepackage{cite}
\usepackage{xcolor}

\begin{document}

\title{{Optimized Resource Allocation for Cloud-Native 6G Networks: Zero-Touch ML Models in Microservices-based VNF Deployments\\}}

\author{\IEEEauthorblockN{Swarna Bindu Chetty\IEEEauthorrefmark{1}, 
                            Avishek Nag\IEEEauthorrefmark{2}, 
                            Ahmed Al-Tahmeesschi\IEEEauthorrefmark{1},
                            Qiao Wang\IEEEauthorrefmark{1},
                            Berk Canberk\IEEEauthorrefmark{3},
                            Johann Marquez-Barja\IEEEauthorrefmark{4},
                            Hamed Ahmadi\IEEEauthorrefmark{1}\\
\IEEEauthorblockA{\IEEEauthorrefmark{1}University of York, UK},\\
\IEEEauthorblockA{\IEEEauthorrefmark{2}University College Dublin, Ireland},\\
\IEEEauthorblockA{\IEEEauthorrefmark{3}Edinburgh Napier University, UK},\\
\IEEEauthorblockA{\IEEEauthorrefmark{4}University of Antwerp, Belgium}
}
}

\maketitle

\begin{abstract}\looseness-1 6G, the next generation of mobile networks, is set to offer even higher data rates, ultra-reliability, and lower latency than 5G. 
New 6G services will increase the load and dynamism of the network. Network Function Virtualization (NFV) aids with this increased load and dynamism by eliminating hardware dependency. It aims to boost the flexibility and scalability of network deployment services by separating network functions from their specific proprietary forms so that they can run as virtual network functions (VNFs) on commodity hardware. It is essential to design an NFV
orchestration and management framework to support these services.
However, deploying bulky monolithic VNFs on the network is difficult, especially when underlying resources are scarce, resulting in ineffective resource management. To address this, microservices-based NFV approaches are proposed. In this approach, monolithic VNFs are decomposed into `micro' VNFs, increasing the likelihood of their successful placement and resulting in more efficient resource management. 
This article discusses the proposed framework for resource allocation for microservices-based services to provide end-to-end \ac{QoS} using the Double Deep Q Learning (DDQL) approach. Furthermore, to enhance this resource allocation approach, we discussed and addressed two crucial sub-problems: the need for a dynamic priority technique and the presence of the low-priority starvation problem. Using the Deep Deterministic Policy Gradient (DDPG) model, an Adaptive Scheduling model is developed that effectively mitigates the starvation problem. Additionally, the impact of incorporating traffic load considerations into deployment and scheduling is thoroughly investigated. 
\end{abstract}


\vspace{-0.1 in}
\section{Introduction}

\IEEEPARstart{T}he evolution of mobile technology from its inception to the current \ac{5G} has continually introduced new opportunities and performance criteria for a wide array of applications. As we transition into the \ac{6G} of mobile networks, we face even more stringent demands for higher reliability, ultra-low latency, and significantly increased data rates~\cite{DTpaper21}. These requirements necessitate tailored network services that can be dynamically deployed and managed to meet various industries' diverse needs while adhering to strict \acp{SLA}.

In traditional, rigid, and monolithic systems, service provisioning, \ac{NS}, often depends on proprietary hardware appliances specifically engineered to optimize particular \ac{NF} tasks. These appliances, designed to meet the objective of \ac{NS}, combine various \acp{NF} to provide seamless support. However, this approach introduces significant limitations, especially when considering deploying 6G applications and services with shorter lifecycles. The tight integration of hardware and software in these systems makes it challenging to modify or customize NSs efficiently, resulting in inflexibility and increased operational and capital expenditures.
It is anticipated that the evolutions of applications will demand a data rate of terabits per second and hundreds of microseconds of latency, significantly increasing the load and dynamism on the network. To support these demands, network infrastructure design, management, and operation have evolved to support these demands, leveraging innovative technologies and architectures. The evolution continues, with microservices-based NFV serving as a continuation of what began with 5G~\cite{10090468}. 

The \ac{NFV} framework transforms traditional hardware-based \acp{NF} like Firewalls and \ac{DPI} into software-based entities known as \acp{VNF}. These VNFs are structural components with clearly defined functional behavior and interfaces that can operate on shared hardware commodities (such as high-volume servers and switches).  This gives \acp{VNF} the freedom, flexibility, and agility to switch between servers in response to dynamic variations in resource demand. The \acp{VNF} are sequentially connected based on the application's functionalities. These sequential connections of VNFs are depicted as directed graphs and are commonly known as \ac{SFC}.
The key advantages of a virtualized infrastructure include flexibility and freedom to deploy new \acp{SFC} and customization of existing \acp{SFC} to meet updated requirements.
In real-world scenarios, the persistent arrival of network services with stringent demands, coupled with a fixed deployment order and continual variations in processing traffic, poses a significant challenge in efficiently distributing resources to individual \acp{VNF}. This challenge, known as the \ac{NFV}-\ac{RA} problem, involves dynamically allocating resources based on real-time demand and ensuring resource elasticity to avoid over-provisioning or under-provisioning for incoming \acp{SFC}. This problem is classified as NP-Hard, indicating its computational complexity.

Effective resource optimization entails efficiently allocating available resources to the services, i.e., maximizing the deployment while minimizing resource wastage. This resource optimization involves various strategies, techniques, and technologies. 
To solve this complexity, one can introduce \ac{AI}/{ML}-based techniques into resource allocation in 6G networks by leveraging advanced algorithms such as deep reinforcement learning, federated learning \cite{YANG202233} and GenAI to optimize network resources in real-time dynamically. These methods allow for efficient, adaptive management of network resources, improving performance and scalability in diverse and complex network environments. \cite{9206115} introduces numerous state-of-the-art resource allocation techniques based on AI/ML to support future 6G networks.

This article presents effective strategies to determine the optimal resource allocation for dynamically arriving network services or \acp{SFC}.
We review the traditional monolithic \ac{VNF}, get familiar with their limitations, and explore the advantages of transitioning towards more efficient microservices-based \acp{SFC}. Subsequently, we emphasize the significance of understanding incoming \acp{SFC} type and their priority within the resource allocation framework. This awareness is critical due to the time-sensitive or mission-critical nature of many of these network services, which often require rapid deployment. Moreover, some of these services are revenue-generating and demand special attention. We also delve into the importance of adaptive scheduling and the role of traffic load in enhancing overall model performance and resource allocation in an end-to-end context.
%
The need to prevent under-provisioning of processing resources during \ac{SFC} deployment is highlighted, along with the transformation of placement into a self-adaptive system capable of managing network traffic fluctuations while considering migration and scaling costs. Additionally, the article explores criteria and algorithms for prioritizing services in scheduling, efficient deployment of mission-critical services, and integrating \ac{ML} for \ac{SFC} placement and resource allocation in dynamic networks. 

\vspace{-0.1 in}
\section{Benefits and Challenges of Microservices}

Traditionally, a monolithic \ac{VNF} is a self-contained, standalone software function operating as a single, cohesive code. Making any changes to this code necessitates access to the entire codebase for building and deploying the updated version, which goes against the agile approach that modern engineers prefer.  This complexity results in large-scale monoliths being challenging to maintain. Finding and patching bugs in these large-scale monoliths requires extensive searches and troublesome fixes within their code base. Moreover, monoliths are subject to `dependency hell,' which results in inconsistent systems that do not compile. Thus, any slight modification to a monolithic application necessitates rebooting the entire application, resulting in significant downtimes and impeding development, testing, and project maintenance.

While \ac{NFV} solves the hardware dependency problem, it introduces new challenges. As mentioned, the monolithic \ac{VNF} approach results in limited flexibility and high rigidity. Finding suitable allocation becomes a complex task for the large-scale monoliths \acp{VNF}, especially when underlying infrastructure resources are scarce. The \ac{NFV} architecture must transform a microservices-based architecture to tackle this challenge.

\emph{Microservices}~\cite{dragoni2017microservices}, 
represent a well-established software design approach. It involves breaking down monolithic systems into smaller, individual self-contained fragments known as `micro.' These micro-fragments comprise loosely coupled software components, offering several benefits, such as simplified maintenance, ease of upgrading, and scaling. In other words, dynamic scaling of \acp{VNF} is possible. Due to their independence, the testing and migration become manageable and allow the re-use of the micro-fragments. 
The microservice concept has proven effective in cloud-based applications such as Netflix and Amazon streaming services. 
Since the telecom networks are becoming `cloud-native,' the microservices-based \ac{NFV} framework promises to serve the future networks well. The virtualization method facilitates the concurrent operation of multiple applications on shared resource infrastructure, which is fundamental to cloud computing.
Hence, resource management is better as it is practical and easy to find placement for lower resource-demanding \acp{VNF} rather than high or large ones. 

\begin{figure*}[ht]
\vspace*{-6pt}
\centering
      \includegraphics[scale= 0.6]{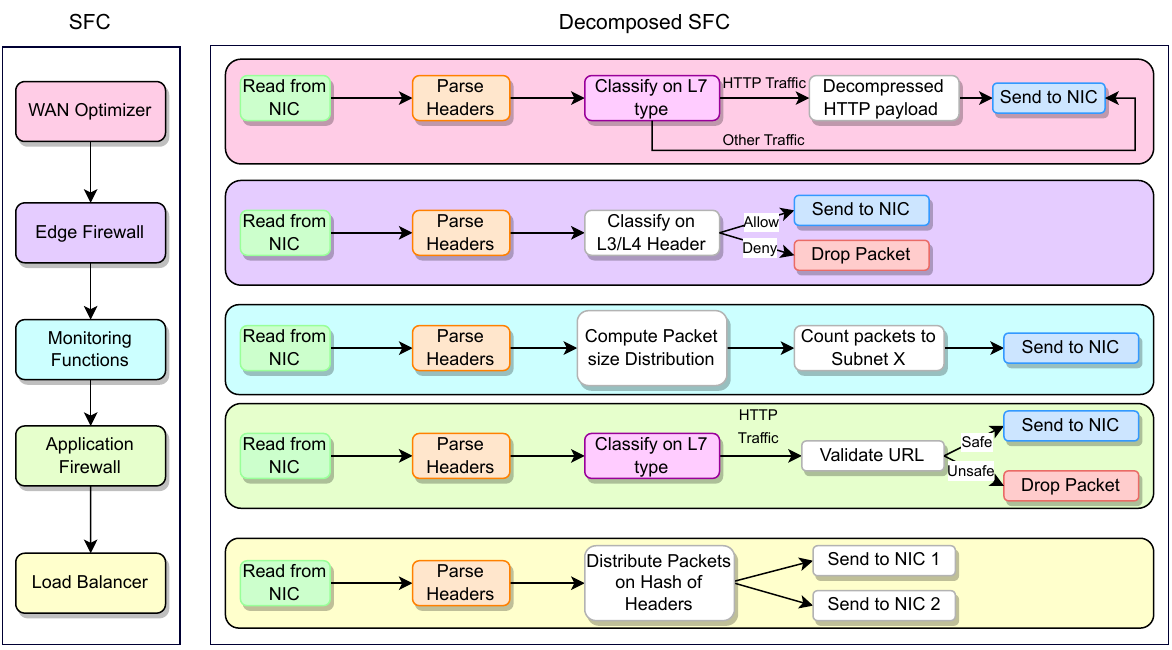} 
      \includegraphics[scale= 0.6]{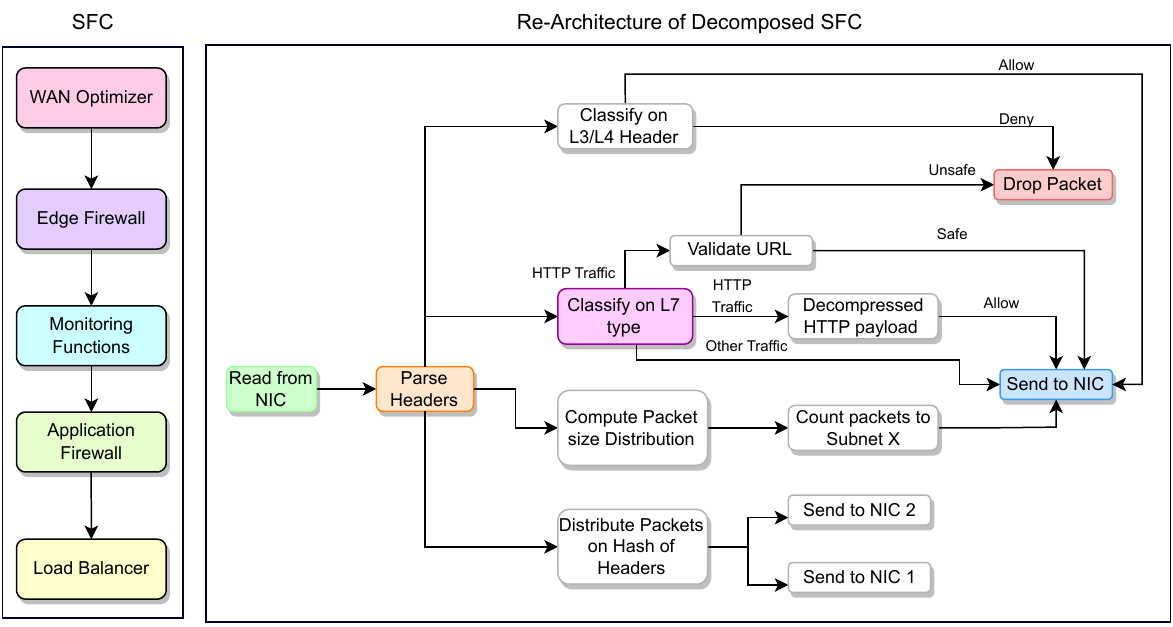}
\caption{(a) An example of Monolith SFC as Decomposed SFC (top), (b) An example of Re-Architecture of Decomposed SFC (bottom)}
\label{chap:4:fig:binpacking1}
\vspace*{-6pt}
\end{figure*}

To better understand how the decomposition of \acp{SFC} looks, Figure \ref{chap:4:fig:binpacking1}(a) illustrates an example of decomposed \ac{SFC} architecture featuring five \acp{VNF}: \ac{WAN} optimizer, Edge Firewall, Monitoring Functions, Application Firewall, and Load Balancer. 
In this setup, packets traverse through each VNF and are processed according to their respective functionalities. The \ac{WAN} Optimizer transmits compressed or decompressed HTTP payloads, including text and images, to optimize bandwidth usage between data centers. To enforce network security policies, the Edge Firewall transmits access control decisions based on layer 2–4 headers, such as IP addresses and TCP/UDP ports. The Monitoring Function transmits metrics and analytics data, including packet size distribution and packet counts for specific IP subnets, to provide visibility into network traffic. The Application Firewall transmits decisions to block or allow HTTP requests based on the presence of SQL injection attacks in URLs, ensuring application layer security. Finally, the Load Balancer transmits traffic distribution commands based on the hash of layer 3–4 headers, efficiently directing traffic to the appropriate backend servers. Within this architecture, each \ac{VNF} is further decomposed into `micro-\acp{VNF}', ensuring the accurate interdependence links between the \acp{VNF} and their corresponding micro-\acp{VNF}. 
These micro-\acp{VNF} are designed to be lightweight and code-based, facilitating effortless deployment onto the network. Moreover, they enable seamless migration and modification, ensuring a hassle-free process. Thus, the \ac{NFV}-\ac{RA} problem gets a different shape with the proposals to decompose monolithic \acp{VNF} into microservices, as the decomposition aids in better resource management.



However, while leveraging these benefits, we encounter new costs and constraints. The deployment and architectural complexity significantly escalate when implementing the microservices concept in the context of the \ac{SFC} embedding problem. In essence, breaking down all \acp{VNF} within an SFC places a greater demand on resources such as bandwidth and latency, as illustrated in Figure~\ref{chap:4:fig:binpacking1}(a). Consequently, an exhaustive decomposition of all VNFs proves to be an inefficient strategy, emphasizing the necessity of pinpointing the optimal decomposition scenario for each SFC while being mindful of the trade-offs involved. In the pursuit of efficient resource allocation, a thoughtful approach is needed. Initially, a model should be established to identify potential \acp{VNF} for decomposition rather than decomposing every \ac{VNF}, aiming to strike a balance in managing the associated overheads. Additionally, instead of assuming a static or pre-defined decomposition, \acp{VNF} should be dynamically decomposed into micro-\acp{VNF} based on a rational approach, such as the availability of physical network resources. A dynamic granularity criterion should be adopted to control the potential \ac{VNF} micro-segmentation level.

Furthermore, several duplicated or redundant micro-VNFs are generated during the decomposition, like `Read from NIC’ and `Parse Header’ as in Figure~\ref{chap:4:fig:binpacking1}(a). Deploying these functions repetitively consumes additional resources. To enhance resource efficiency, one practical approach involves adopting a re-architecting strategy to eliminate such redundancy in micro-VNFs. Subsequently, the microservices-based SFC architecture can be revised or customized accordingly. 
Figure~\ref{chap:4:fig:binpacking1}(b) illustrates a comprehensive re-architecture of the decomposed SFC, demonstrating that redundant micro-VNFs (`Read from NIC,' `Parse Header') are deployed only once on the network based on demand. As per the VNF requests, the flow has been activated or utilized. Importantly, this approach does not disrupt any internal flows. Thus, this optimization effectively reduces the overhead introduced during the decomposition process, ultimately enhancing overall resource efficiency.

\vspace{-0.1 in}
\section{Prioritization \& Scheduling Style}
\subsection{The Importance of Embracing Dynamic Prioritization}
Prioritization is the methodology of assessing tasks' relative significance, which helps determine the critical task and should be completed first.  Applying the same concept, prioritization of network services entails identifying the most crucial services, allocating resources, and altering the preferences accordingly.One standard method of prioritizing network services is by understanding \ac{QoS}; depending on business requirements (i.e., applications) or analyzing the incoming traffic load based on the policy, the network administrators define different priority levels.
Not all SFCs are identical or have the same priority level in the real world. Consequently, a one-size-fits-all approach does not suffice. When relying on traditional methods like \ac{FIFO} or \ac{FCFS}~\cite{9472805, 10.1007/s11227-021-03784-7}, there is a tendency to overlook the importance of distinguishing and prioritizing emergency \acp{SFC} over non-emergency ones. This oversight can result in the rejection of highly critical \acp{SFC} as resources might be exhausted or the waiting time for the services has expired. Thus, prioritization of arriving \ac{SFC} is essential for better resource management. The existing literature often used a prevalent approach involving a rigid priority or manual classification. It is reasonable for earlier systems because the network services were distinct, less complicated, and rarely evolving. However, this inflexible approach becomes ineffective as we look ahead to dynamically evolving \acp{SFC} with time-varying resource requirements. Therefore, transitioning from manual (offline) prioritization to dynamic (online) prioritization is necessary to support these incoming \ac{SFC}. In the standard and simplistic view, the services are categorized into two cardinalities (i.e., \ac{PRM} and \ac{BE})~\cite{cappanera2019vnf, mohamad2020psvshare}. 
The work of Francesco \emph{et al.}~\cite{malandrino2019reducing} and Maryam \emph{et al.}~\cite{jalalitabar2016service} 
demonstrated the mapping of the services depending on the priority of \ac{VNF}, where this priority is assigned randomly. Many researchers~\cite{8540003, 9045275} 
performed a multi-class prioritization, like Rongpeng \emph{et al.}~\cite {8540003} described three levels of priority (category A, B, and C). 
Most researchers considered services that had already been prioritized without defining any analytical criteria. The prioritization is performed offline, which requires extra time and resources. This raises questions like, `Will this rigid approach be compatible for next-generation services, which diminish the value of services relative to one another within a class?' 

A more realistic way of approaching this problem is presented in~\cite{chetty2023dynamic}, 
where an intelligent (\ac{ML}-based) dynamic prioritization for online service is established, which allocates two-tier priority to the online services.

\vspace{-0.1 in}
\subsection{Scheduling: A Must-Have}
In the works~\cite{cappanera2019vnf, mohamad2020psvshare}, there is a consistent emphasis on prioritizing \ac{PRM} services over \ac{BE} services. Likewise, in the studies~\cite{8540003, 9045275}, the SFC deployment is based on a hierarchy of service priority from high to low. While being beneficial for priority services, this approach unfortunately leads to significant service deprivation for low-priority services.

A successful deployment strategy considers service prioritization and implements intelligent scheduling of incoming services based on diverse factors. The scheduler should adhere to a defined policy for efficiently deploying services onto the network. As noted earlier, the FIFO approach presents limitations that can be mitigated by adopting a priority-based deployment strategy. However, upon closer examination, this strategy reveals fresh challenges, specifically concerning `starvation for lower-priority services.'
This issue, which researchers have ignored, introduces a substantial bias within the system, potentially resulting in an unfair user experience. Thus, aiming to strike a balance between prioritizing critical \acp{SFC} and ensuring fair treatment for lower-priority services is essential. 
A biased scheduling system prompts important questions such as, ``Can the priority attribute be considered the most effective factor in scheduling decision-making?'' and ``Would the decision be more optimal by taking into account additional factors, such as service waiting time or reliability?'' The following section tackles these questions through a detailed case study.

\vspace{-0.1 in}
\section{Case Study: Dynamic Strategies for Next-Gen Network Efficiency on Real Networks—Prioritization and Scheduling in 6G}
In our pursuit of a solution, we introduce a \ac{DyPr} model and an \ac{AdSch} model to prioritize service on-the-fly and at-that-point scheduling systems rather than the traditional rigid one. Additionally, we delve into the effects of traffic load to establish an efficient resource allocation system. 

\vspace{-0.1 in}
\subsection{\ac{ML}-Based Dynamic Prioritization (DyPr)}
Our model\ac{ML}-Based \ac{DyPr} takes a nuanced approach by expressing each service's priority through macro-class and micro-class levels. The micro-class priority indicates status within a specific class, while the macro-class priority categorizes the broader class, ranked from 0.0 to 1.0. A higher value signifies greater priority. Formulated as an online-ridge regression approach, \ac{ML}-Based \ac{DyPr} performs L2 regularization to reduce multicollinearity among decision-making parameters. This model includes an observatory phase before the learning phase to handle unseen services without pre-existing datasets. During the observatory phase, the model gathers data about online services in a memory buffer until sufficient transitions are collected. Initially, priorities are uniformly allocated to construct training datasets, creating an `Online-RR' learning model. Once enough data is gathered, the learning phase begins, and the transitions are selected randomly to prevent over-fitting. The model's accuracy is periodically checked, shifting from `Train' to `Predict' mode once the desired accuracy is achieved. This zero-touch cognitive system learns from independent variables like threshold jitters, delay, and packet loss to determine appropriate priority levels. This model identifies hidden patterns, understands evolving service types, and allocates appropriate priority levels based on \ac{QoS} and flow types. It provides granular information on the relative importance of services, proving highly advantageous for time-sensitive or mission-critical applications during the adaptive scheduling phase. Consequently, the \ac{ML}-Based \ac{DyPr} model is considered a regression model in our research, as it captures sensitive information about the services. In-depth information about the model is provided in \cite{10494765}.

\vspace{-0.1 in}
\subsection{Adaptive Scheduling (AdSch)}
We assume some services (of any priority) can be more frequently requested and revenue-generating; such services must be classified as High-Demand and must be considered during scheduling and deployment. To achieve this balance, we propose strategies that optimize resource allocation while preventing unfairness in the system.
In our pursuit of an impartial scheduling system, we have proposed an intelligent \ac{AdSch} module utilizing the \ac{DDPG} approach~\cite{chetty2023dynamic}. This method considers multiple factors and aims to make optimal decisions. For instance, when comparing a high-priority \ac{SFC} A with a higher waiting threshold to a lower-priority \ac{SFC} B about to expire soon, the traditional model would select \ac{SFC} A, disregarding \ac{SFC} B's impending expiration, potentially affecting the \ac{QoE}. However, our model, accounting for multiple factors, prioritizes \ac{SFC} B over \ac{SFC} A. It understands that allowing \ac{SFC} A to wait a bit longer in the queue will not adversely impact it, thus optimizing the scheduling process. A \ac{DDPG}, which is an actor-critic \ac{RL} model, is trained to recognize 'Beneficial and Starving' services based on three factors: \ac{TWT}, Reliability, and Priority of the service, which act as the state-space. Based on these factors, the DDPG model assigns a rank between 0 and 1 to each requested service, indicating the service's significance. A higher rank signifies a greater necessity to deploy the service, leading to a rank-based scheduling method. An appropriate reward function is essential to train the DDPG model. The reward function comprises Beneficial-cost and Starvation-cost, balancing high-priority, starvation, and high-reliability services. In-depth information about the model is provided in \cite{chetty2023dynamic}.
%
%
The proposed method initially allocated resources to critical and beneficial services before assigning them to other services. This approach significantly enhances resource allocation management, ultimately benefiting the overall system efficiency.

\vspace{-0.1 in}
\subsection{Traffic Load-Based Service Classifier}
In addition to examining the advantages of \ac{DyPr} and \ac{AdSch}, our research delved into understanding the influence of traffic load resulting from various services on resource management. Furthermore, we aimed to minimize any potential elastic cost, specifically the migration and scaling expenses that could arise from these services.

In networking, the volume of data or information from the source to the destination node (within a network) is often called traffic or workload. Various types of network traffic, such as data packets, voice calls, and video streams, are exchanged between interconnected devices based on the applications and services in use. The various types of network traffic contribute to the network load, and this load undergoes fluctuations influenced by factors such as the time of day, typical network traffic patterns, and the specific kinds of applications or services being utilized. Traffic complexity intensifies, especially within denser heterogeneous architectures like those in \ac{5G} and beyond networks. This highlights the importance of understanding network traffic and its implications for service deployment and resource management.

In \ac{6G}, considering the dynamic time-varying traffic, the required processing resources for a \ac{VNF} are expected to fluctuate based on traffic arrival statistics and the \ac{QoS} requirements. 
The processing load for each service changes regularly, influenced by variations in the number of users or clients requesting the service. This consequently affects the generated traffic load.
A deep understanding of this time-varying network traffic is imperative to maintain a reliable and efficient network operation. 
Network administrators require a deep understanding of the network load to make knowledgeable decisions on capacity planning and resource allocation. Their role is vital in optimizing network performance by ensuring adequate resources to effectively meet user demands. 
 Aligning \ac{VNF} deployment with traffic load helps prevent over-utilization and under-utilization of processing and transmission resources. Thus,  it is advantageous to consider traffic characteristics for \ac{VNF} deployment. 

To address this, we propose a supervised classifier for binary classification of all incoming \acp{SFC}: \ac{HD} and \ac{NHD}. However, a key challenge is the constantly changing network load and the unavailability of labeled data required to train supervised models.
We started by clustering the collected service data into highly demanding (popular) and less demanding services among users or clients to overcome this challenge. Subsequently, we applied unsupervised models such as K-means, Agglomerative clustering, \ac{BIRCH}, and \ac{GM} to convert unlabelled data into labeled data. This labeled data became instrumental in training our supervised learning models. Post-training, the model proficiently classified online services as either \ac{HD} and \ac{NHD}. \ac{SFC} clustering and classification consider both \ac{QoS} and cumulative load over an \ac{SFC}'s lifespan, including packet volume in defined time frames. For more information on traffic generation and detailed performance analysis of various unsupervised models, please refer to~\cite{10494765}.

Now, the labeled incoming request needs to be scheduled. The results are fed into the \ac{AdSch} model, generating a scheduling queue. Based on this queue, one-by-one \ac{SFC} are deployed, allowing us to analyze the impact of traffic load on them.

\vspace{-0.1 in}
\subsection{Overview of Microservices-based SFC deployment}
The primary objective is to identify potential VNFs, referred to as candidate VNFs, for decomposition that would optimize network resource consumption. To achieve this goal, each VNF undergoes verification using a decomposition identification module to determine its suitability as a prospective candidate for dynamic decomposition. One of the criteria considered is when the model cannot identify a suitable or satisfactory substrate node for deployment, i.e., when the selected node's available resource does not satisfy the demanded resources. After identifying potential VNFs for decomposition, the next consideration is determining the optimal level of decomposition. We propose a dynamic granularity criterion represented by the Granularity Index (GI). The GI is based on the VNF's requested processing resource (CPU core) and the Network Availability Index (NAI), which signifies the ratio of currently available processing resources to the total initial capacity in the network. The NAI ranges from 0 to 1, offering insights into possible micro-segmentation combinations for a VNF. After the decomposition, the identification model verifies the presence of the newly created micro-VNFs in the network. If successful, it retrieves essential information, such as the deployed location (substrate node), from the repository. This triggers the re-architecture model, connecting the existing micro-VNF with the previous one to construct a microservices-based SFC aligning with the promised QoS requirements. If unsuccessful, the newly developed micro-VNF resource requirements are provided to the DDQL agent to identify a suitable substrate node for deployment, and these details are cataloged for future reference.

\begin{figure}[ht]
    \vspace*{-6pt}
    \centering
    \includegraphics[width=0.9\linewidth]{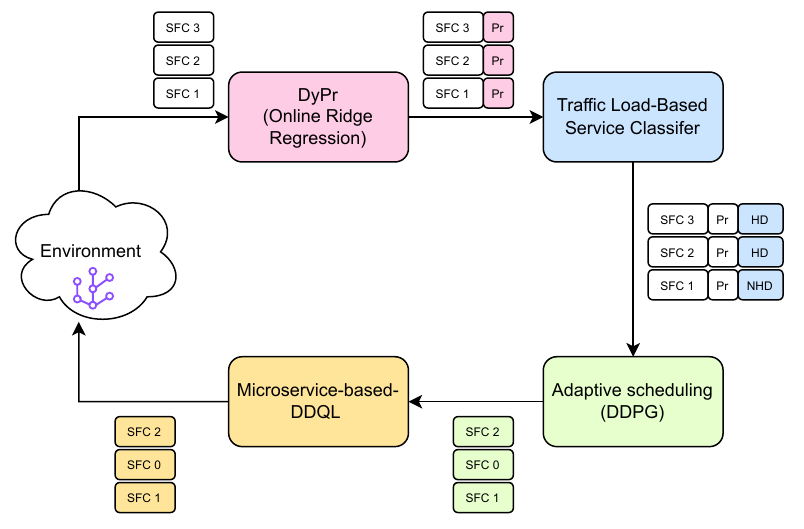}
    \caption{Overview of the Proposed Model}
    \label{fig:enter-label}
\end{figure}

Finally, a well-planned and optimized deployment strategy is needed, ensuring that network resources are allocated optimally, aligned with the demands of the network, and maximizing overall performance. Thus, we focus on allocating resources to the services, i.e., solving the NFV-RA problem using the advanced \ac{RL} technology: the \ac{DDQL} approach to maximize the deployment of \acp{SFC} onto the substrate network while following defined constraints, as mentioned in \cite{10494765}.

Our proposed framework conceptualized the physical topology as an environment characterized by many high-capacity servers. Each \ac{SFC} and its unique resource demands were meticulously outlined as the state space ($S$). The action space ($A$) was defined by the count of physical nodes or servers constituting the topology. This precise representation was crucial in conducting our analyses and simulations. In our study, we developed two primary reward functions: Global and Local reward functions. The global reward function is assigned to the agent upon successfully deploying the \ac{SFC} along with all its associated \acp{VNF}. On the other hand, the local reward function is specifically designed for each \ac{VNF}, considering factors such as the quality of the selected node for placement, priority, reliability, successful placement, and traffic load. These reward functions are pivotal in reinforcing the learning process of our model. A detailed formulation of the reward functions, state space, and action space can be found in~\cite{10494765}.

To give a comprehensive overview, the services that arrive are first prioritized based on their QoS and other attributes using the DrPy model. These prioritized services are then passed to the servicer classifier, determining whether they are classified as HD or NHD. With this classification information, the Adaptive Scheduling model reorganizes the queue for deployment. Finally, the reordered queue is processed by the Microservice-based DDQL model for optimal service placement, as shown in Figure~\ref{fig:enter-label}. DDPG is selected for its ability to handle continuous action spaces, making it suitable for adaptive scheduling where decisions involve continuous variables. DDQN is chosen for its robustness in managing discrete action spaces typically encountered in resource allocation tasks, thereby reducing the likelihood of overestimation biases inherent in standard Q-learning methods.

\vspace{-0.1 in}
\subsection{Simulation Parameters}
The simulation results provide a detailed understanding of the effects of traffic-aware \ac{VNF} deployment using the \ac{DDQL} and \ac{AdSch} models. We conducted a {comprehensive performance analysis of the models on two distinct network topologies: Netrail (7 nodes, 10 links) and BtEurope (24 nodes, 37 links). The evaluation confined a spectrum of substrate nodal and link capacities from highly available resource topologies to those more prone to exhaustion.  Due to the similar behavior exhibited by both topologies, we only represent Netrail. The online services were generated utilizing the Erdős–Rényi model, varying in structural complexity and resource requirements,i.e., each SFC is composed of variable VNFs and requires distinctive SLA. The arrival of SFCs follows a Poisson distribution. Each QoS parameter for these SFCs, such as delay, jitter, packet loss, and other relevant metrics, is generated using a Gaussian distribution with different means and standard deviations, as mentioned in \cite{10494765}. 
Each simulation contained 2000 episodes, and within each episode, a maximum of 100 services were expected. The machine learning models were implemented using Python and the PyTorch library. All simulations were executed on hardware featuring an Intel Core i7 processor with 64 GB RAM.

\vspace{-0.1 in}
\subsection{Outcomes}
We explored the service placement performance of our proposed framework (i.e., \ac{DDQL} + online-Ridge Regression + \ac{DDPG} + traffic generator) across various priority service types, using a moving average of normalized Service Acceptance Rate (SAR) over the last 100 episodes for detailed analysis, as shown in Figure~\ref{fig1}. 

Following a rigorous analysis, we selected two optimal models: Agglomerative combined with Decision Tree and K-means combined with Logistic Regression for clustering and classifying services. These models were complemented by the \ac{DDQL} deploying model and \ac{AdSch} model. To gauge their effectiveness, we compared the performance of these models with the \ac{DDQL} traffic-aware-less (or as stated without traffic Load ) scenario. 
The K-means with Logistic Regression (Kmeans+LR) scenario did not display optimal performance in our evaluations. Even after numerous episodes, it failed to surpass the performance of the conventional system. Contrastingly, starting from episode 800, the Agglomerative combined with Decision Tree (Aggo+DT) models showcased superior performance by consistently outperforming others in service deployment. This model demonstrated a high level of proficiency in learning the objective function, maximizing the deployment rate by efficient resource allocation. 

\begin{figure}[ht]
    \vspace*{-6pt}
    \centerline{\includegraphics[width=18.5pc]{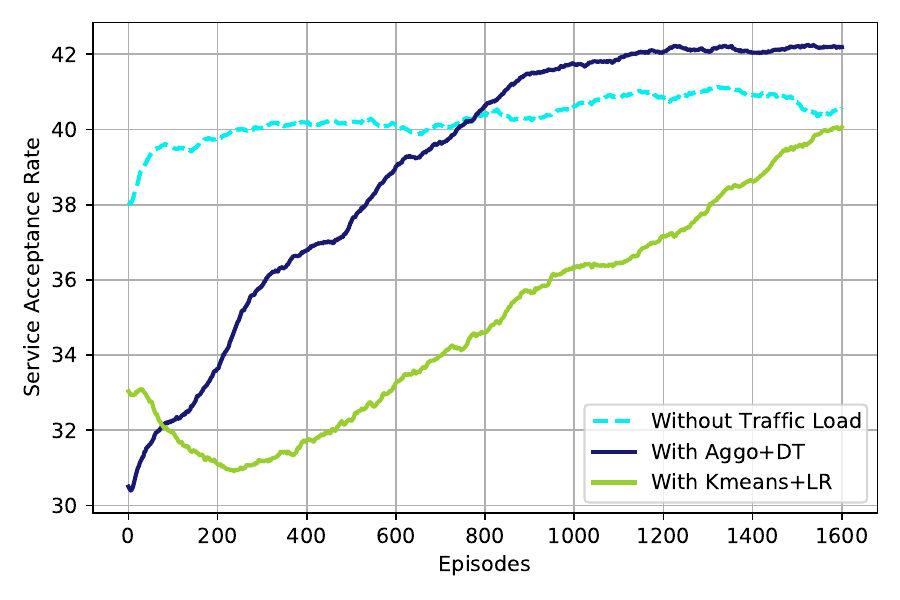}}
    \caption{Proposed Model's Performance for All Priority Services in Terms of Deployment}
    \vspace*{-6pt}
    \label{fig1}
\end{figure}

Insights into the models’ performance regarding remaining resources in the network are detailed in Figure~\ref{fig2}. The visualization emphasizes how the Aggo+DT models adeptly learn the optimal approach to deploy incoming services efficiently. On average, each service required 15 or more CPU core resources. Furthermore, the Aggo+DT model demonstrated a balanced resource utilization approach, avoiding excessive and inadequate network usage and ensuring ample resources for other essential functionalities. Indeed, the results indicate that the model was trapped in a sub-optimal solution without traffic awareness, leading to an under-utilized network. This emphasizes integrating traffic awareness to achieve efficient resource allocation and maximize network utilization. 

\begin{figure}[ht]
    \vspace*{-6pt}
    \centerline{\includegraphics[width=18.5pc]{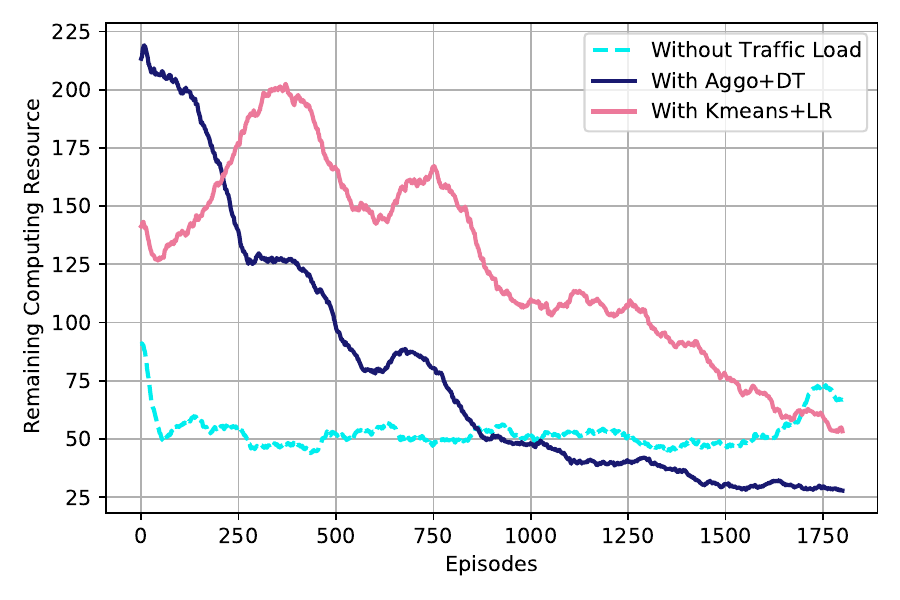}}
    \caption{Proposed Model's Performance for All Priority Services in terms of Remainder Computing Resource}
    \vspace*{-6pt}
    \label{fig2}
\end{figure}

Our analysis, showcased through Figures \ref{fig3} and \ref{fig4} offer a thorough analysis of how our proposed framework effectively mitigates the starvation problem while ensuring uninterrupted service for high-priority tasks. These visualizations delve into the core of our framework's ability to strike a balance, preventing resource starvation for lower-priority services while maintaining the smooth functioning of critical high-priority services. Our analysis compared the overall performance of deploying models (\ac{DDQL}) with and without traffic awareness. Figure \ref{fig3} illustrates that with traffic awareness, the \ac{DDQL} and \ac{AdSch} models prioritized low-priority services for deployment, effectively addressing the starvation problem. On the other hand, the Aggo+DT model could deploy twice as many services without traffic awareness. 
Furthermore, in a comparative analysis between the Aggo+DT and Kmeans+LR models, the former showcased faster learning and convergence during training. This indicates the superior learning efficiency of Aggo+DT compared to Kmeans+LR.
The analysis raises a pertinent question: ``Do high-priority services experience any negative impact?'' Figure \ref{fig4} serves as a conclusive response. The Aggo+DT model learns to prioritize high-priority services, ultimately surpassing the model's performance without traffic awareness. Conversely, the Kmeans+LR model encounters challenges in effectively understanding and managing service, struggling to optimize performance compared to Aggo+DT.

In conclusion, our proposed framework, especially with Aggo+DT traffic awareness, showcased superior outcomes in resource allocation. Notably, it effectively addressed critical sub-problems, including the starvation of low-priority services and potential negative impacts on high-priority services, ensuring the network's optimal utilization without overuse or underuse. These findings highlight the framework's efficiency in achieving optimal resource allocation while guaranteeing the seamless operation of vital services.

\begin{figure}
    \vspace*{-6pt}
    \centerline{\includegraphics[width=18.5pc]{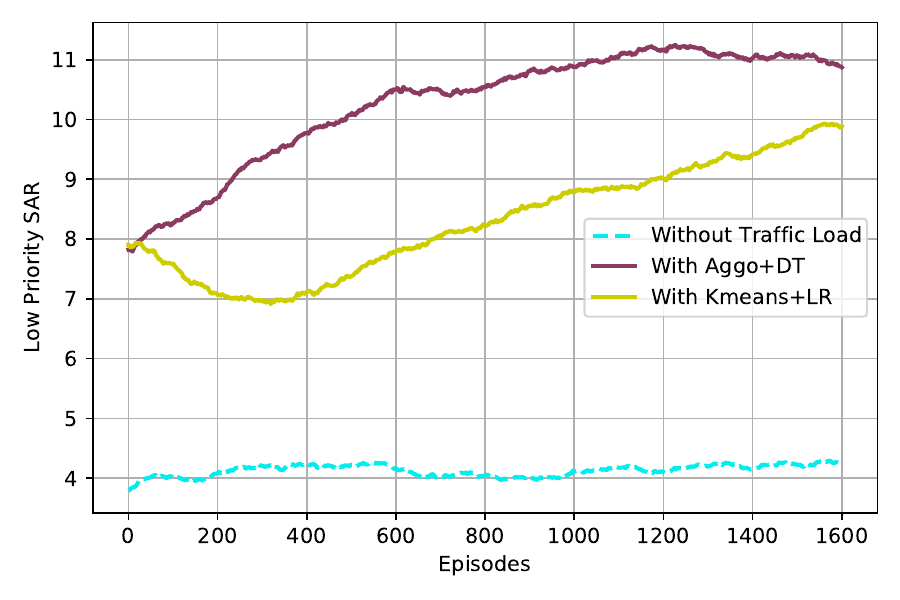}}
    \caption{Proposed Model's Performance for Low Priority Services in Terms of Deployment}
    \vspace*{-6pt}
    \label{fig3}
\end{figure}

\begin{figure}
    \vspace*{-6pt}
    \centerline{\includegraphics[width=18.5pc]{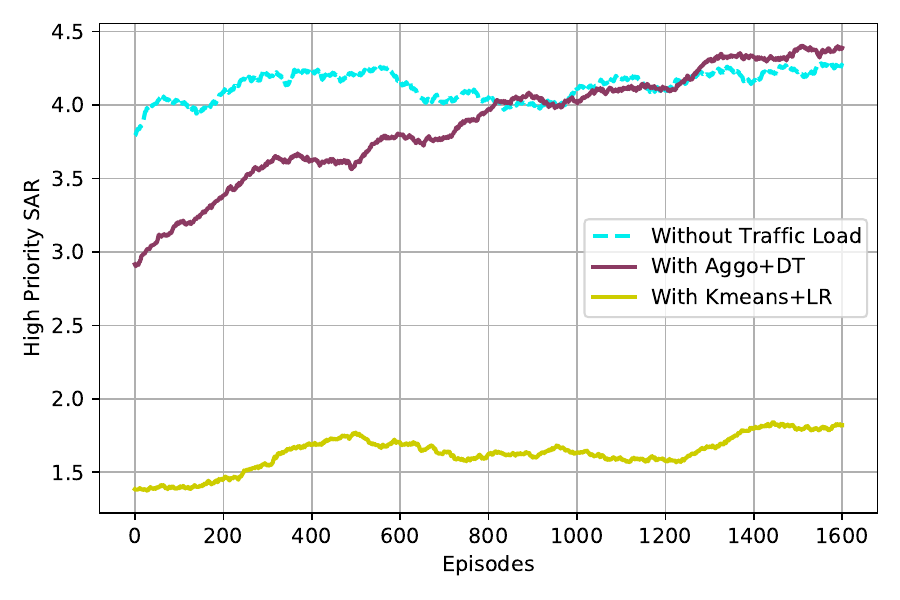}}
    \caption{Proposed Model's Performance for High Priority Services in Terms of Deployment}
    \vspace*{-6pt}
    \label{fig4}
\end{figure}

\vspace{-0.1 in}
\section{Conclusion}

In the domain of \ac{6G} networks, various network services with stringent requirements across different industries are anticipated. This article enhances support for these evolving network services within future systems. Conventional approaches to deploying incoming \acp{SFC}, as highlighted in the literature, fall short and pose significant challenges. Hence, integrating \ac{ML}-assisted systems is essential for tackling critical challenges comprehensively. This includes optimizing resource allocation for online services, mitigating low-priority starvation problems without adversely affecting high-priority services, and strategically incorporating traffic awareness into scheduling and deployment processes. By doing so, we pave the way for more efficient and adaptive resource allocation in the network, poised to meet the dynamic demands of evolving network services while ensuring a seamless and unbiased user experience.
Future research should focus on scalable solutions, improved real-time adaptation, accurate traffic predictions, and diverse evaluations. Additionally, exploring network slicing, advanced ML models for resource allocation, and compatibility with ORAN architecture will be crucial. Addressing these areas will enhance the robustness and adaptability of the framework for optimized resource allocation in cloud-native 6G networks.

\vspace{-0.1 in}
\begin{acronym} 
\acro{5G}{the fifth generation of mobile networks}
\acro{6G}{sixth generation of mobile networks}
\acro{ACO}{Ant Colony Optimization}
\acro{AI}{Artificial Intelligence}
\acro{AR}{Augmented Reality}
\acro{ANN}{Artificial Neural Network}
\acro{AdSch}{`Adaptive Scheduling’}
\acro{BB}{Base Band}
\acro{BBU}{Base Band Unit}
\acro{BER}{Bit Error Rate}
\acro{BS}{Base Station}
\acro{BW}{Bandwidth}
\acro{BIRCH}{Balanced Iterative Reducing and Clustering using Hierarchies}
\acro{BE}{Best-Effort}
\acro{CC}{Chain Composition}
\acro{C-RAN}{Cloud Radio Access Networks}
\acro{CAPEX}{Capital Expenditure}
\acro{CoMP}{Coordinated Multipoint}
\acro{COTS}{Commercial-Off-The-Shelf}
\acro{CR}{Cognitive Radio}
\acro{COC}{Computation Oriented Communications}
\acro{CAeC}{Contextually Agile eMBB Communications}
\acro{CBQ}{Class-Based Queueing}
\acro{D2D}{Device-to-Device}
\acro{DA}{Digital Avatar}
\acro{DAC}{Digital-to-Analog Converter}
\acro{DAS}{Distributed Antenna Systems}
\acro{DBA}{Dynamic Bandwidth Allocation}
\acro{DNN}{Deep Neural Network}
\acro{DC}{Duty Cycle}
\acro{DyPr}{`Dynamic Prioritization'}
\acro{DL}{Deep Learning}
\acro{DSA}{Dynamic Spectrum Access}
\acro{DT}{Digital Twin}
\acro{DRL}{Deep Reinforcement Learning}
\acro{DQL}{Deep Q Learning}
\acro{DDQL}{Double Deep Q Learning}
\acro{DTs}{Decision Tress}
\acro{DDPG}{Deep Deterministic Policy Gradient}
\acro{DPI}{Deep Packet Inspection}
\acro{$E^2D^2PG$}{Enhanced Exploration Deep Deterministic Policy}
\acro{ER}{Erdős-Rényi}
\acro{EUB}{Expected Upper Bound}
\acro{EDuRLLC}{Event Defined uRLLC}
\acro{EVNFP}{Elastic Virtual Network Function Placement}
\acro{eMBB}{enhanced Mobile Broadband}
\acro{FBMC}{Filterbank Multicarrier}
\acro{FEC}{Forward Error Correction}
\acro{FG} {Forwarding Graph}
\acro{FGE}{FG Embedding}
\acro{FIFO}{First-in-First-out}
\acro{FCFS}{First-Come-First-Served}
\acro{FFR}{Fractional Frequency Reuse}
\acro{FSO}{Free Space Optics}
\acro{fBm}{Fractional Brownian motion}
\acro{GA}{Genetic Algorithms}
\acro{GI}{Granularity Index}
\acro{GM}{Gaussian Mixture}
\acro{HAP}{High Altitude Platform}
\acro{HD}{High-Demand}
\acro{HL}{Higher Layer}
\acro{HARQ}{Hybrid-Automatic Repeat Request}
\acro{IoE}{Internet of Everything}
\acro{IoT}{Internet of Things}
\acro{ILP}{Integer Linear Program}
\acro{KPI}{Key Performance Indicator}
\acro{KNN}{K-Nearest Neighbour}
\acro{LAN}{Local Area Network}
\acro{LAP}{Low Altitude Platform}
\acro{LL}{Lower Layer}
\acro{LR}{Logistic Regression}
\acro{LOS}{Line of Sight}
\acro{LTE}{Long Term Evolution}
\acro{LTE-A}{Long Term Evolution Advanced}
\acro{LRD}{Long-Range Dependence}
\acro{LFGL}{Least-First-Greatest-Last}
\acro{MAC}{Medium Access Control}
\acro{MAP}{Medium Altitude Platform}
\acro{MIMO}{Multiple Input Multiple Output}
\acro{ML}{Machine Learning}
\acro{MME}{Mobility Management Entity}
\acro{mmWave}{millimeter Wave}
\acro{MNO}{Mobile Network Operator}
\acro{MR}{Mixed Reality}
\acro{MILP}{Mixed-Integer Linear Program}
\acro{MDP}{Markov Decision Process}
\acro{mMTC}{massive Machine Type Communications}
\acro{NAI}{Network Availability Index}
\acro{NASA}{National Aeronautics and Space Administration}
\acro{NAT}{Network Address Translation}
\acro{NHD}{Not-so-High-Demand}
\acro{NN}{Neural Network}
\acro{NF}{Network Function}
\acro{NFP}{Network Flying Platform}
\acro{NTN}{Non-terrestrial networks}
\acro{NFV}{Network Function Virtualization}
\acro{NS}{Network Service}
\acro{OFDM}{Orthogonal Frequency Division Multiplexing}
\acro{OSA}{Opportunistic Spectrum Access}
\acro{OPEX}{Operational Expenditure}
\acro{PAM}{Pulse Amplitude Modulation}
\acro{PAPR}{Peak-to-Average Power Ratio}
\acro{PGW}{Packet Gateway}
\acro{PHY}{physical layer}
\acro{PSO}{Particle Swarm Optimization}
\acro{PT}{Physical Twin}
\acro{PU}{Primary User}
\acro{PRM}{Premium}
\acro{QAM}{Quadrature Amplitude Modulation}
\acro{QoE}{Quality of Experience}
\acro{QoS}{Quality of Service}
\acro{QPSK}{Quadrature Phase Shift Keying}
\acro{QL}{Q-Learning}
\acro{RA}{Resource Allocation}
\acro{RF}{Radio Frequency}
\acro{RN}{Remote Node}
\acro{RRH}{Remote Radio Head}
\acro{RRC}{Radio Resource Control}
\acro{RRU}{Remote Radio Unit}
\acro{RL}{Reinforcement Learning}
\acro{RR}{Ridge Regression}
\acro{SCH}{Scheduling}
\acro{SU}{Secondary User}
\acro{SCBS}{Small Cell Base Station}
\acro{SDN}{Software Defined Network}
\acro{SFC}{Service Function Chaining}
\acro{SLA}{Service Level Agreement}
\acro{SNR}{Signal-to-Noise Ratio}
\acro{SON}{Self-Organising Network}
\acro{SAR}{Service Acceptance Rate}
\acro{SVM}{Support Vector Machine}
\acro{SLFL}{Simple Lazy Facility Location}
\acro{TDD}{Time Division Duplex}
\acro{TD-LTE}{Time Division LTE}
\acro{TDM}{Time Division Multiplexing}
\acro{TDMA}{Time Division Multiple Access}
\acro{TWT}{Threshold Waiting Time}
\acro{UE}{User Equipment}
\acro{UAV}{Unmanned Aerial Vehicle}
\acro{URLLC}{Ultra-Reliable Low Latency Communications}
\acro{USRP}{Universal Software Radio Platform}
\acro{VL}{Virtual Link}
\acro{VNF}{Virtual Network Function}
\acro{VNF-FG}{VNF-Forwarding Graph}
\acro{VNF-FGE}{VNF-FG Embedding}
\acro{VM}{Virtual Machine}
\acro{VR}{Virtual Reality}
\acro{WAN}{Wide Area Network}
\acro{WFQ}{Weighted Fair Queuing}
\acro{XAI}{Explainable Artificial Intelligence}
\acro{XR}{Extended Reality}
\end{acronym}
\bibliographystyle{IEEEtran}

\bibliography{reference.bib}
\end{document}